\documentclass[%
 reprint,
 superscriptaddress,
 amsmath,amssymb,
 prl,
floatfix,
]{revtex4-2}

\usepackage{graphicx}% Include figure files
\usepackage{dcolumn}% Align table columns on decimal point
\usepackage{bm}% bold math
\usepackage{color,soul}
\usepackage{ulem} 
\usepackage{hyperref}%
\DeclareUnicodeCharacter{2212}{-}
\usepackage{float}
\usepackage{ulem}
\begin{document}

\title{Interplay between atomic fluctuations and charge density waves in La$_{2-x}$Sr$_{x}$CuO$_{4}$}

\author{L. Shen}
\email{lingjia.shen@sljus.lu.se}
 \affiliation{Division of Synchrotron Radiation Research, Lund University, SE-22100 Lund, Sweden}
%  \email{lingjia.shen@sljus.lu.se}
 \affiliation{Stanford Institute for Materials and Energy Sciences, Stanford University and SLAC National Accelerator Laboratory,
Menlo Park, California 94025, USA}
 \affiliation{Linac Coherent Light Source, SLAC National Accelerator Laboratory, Menlo Park, California 94025, USA}
 
 \author{V. Esposito}
 \affiliation{Stanford Institute for Materials and Energy Sciences, Stanford University and SLAC National Accelerator Laboratory,
Menlo Park, California 94025, USA}
 \affiliation{Linac Coherent Light Source, SLAC National Accelerator Laboratory, Menlo Park, California 94025, USA}
 
 \author{N. G. Burdet}
 \affiliation{Stanford Institute for Materials and Energy Sciences, Stanford University and SLAC National Accelerator Laboratory,
Menlo Park, California 94025, USA}
 \affiliation{Linac Coherent Light Source, SLAC National Accelerator Laboratory, Menlo Park, California 94025, USA} 
 
 \author{M. Zhu}
 \affiliation{H.H. Wills Physics Laboratory, University of Bristol, Bristol BS8 1TL, United Kingdom}
 
 \author{A. N. Petsch}
 \affiliation{H.H. Wills Physics Laboratory, University of Bristol, Bristol BS8 1TL, United Kingdom}
 \author{T. P. Croft}
 \affiliation{H.H. Wills Physics Laboratory, University of Bristol, Bristol BS8 1TL, United Kingdom}

 \author{S. P. Collins}
 \affiliation{Diamond Light Source Ltd., Harwell Science and Innovation Campus, Didcot, Oxfordshire OX11 0DE, United Kingdom}
  
 \author{Z. Ren}
 \affiliation{Deutsches Elektronen-Synchrotron (DESY), Notkestraße 85, 22607 Hamburg, Germany}
 
 \author{F. Westermeier}
 \affiliation{Deutsches Elektronen-Synchrotron (DESY), Notkestraße 85, 22607 Hamburg, Germany}
 
 \author{M. Sprung}
 \affiliation{Deutsches Elektronen-Synchrotron (DESY), Notkestraße 85, 22607 Hamburg, Germany}

 \author{S. M. Hayden}
 \email{S.Hayden@bristol.ac.uk}
 \affiliation{H.H. Wills Physics Laboratory, University of Bristol, Bristol BS8 1TL, United Kingdom}
 
 \author{J. J. Turner}
 \email{joshuat@slac.stanford.edu}
 \affiliation{Stanford Institute for Materials and Energy Sciences, Stanford University and SLAC National Accelerator Laboratory,
Menlo Park, California 94025, USA}
 \affiliation{Linac Coherent Light Source, SLAC National Accelerator Laboratory, Menlo Park, California 94025, USA}
 
\author{E. Blackburn}
\email{elizabeth.blackburn@sljus.lu.se}
\affiliation{Division of Synchrotron Radiation Research, Lund University, SE-22100 Lund, Sweden}

\begin{abstract}
In the cuprate superconductors, the spatial coherence of the charge density wave (CDW) state grows rapidly below a characteristic temperature $T_\mathrm{CDW}$, the nature of which is debated. We have combined a set of x-ray scattering techniques to study La$_{1.88}$Sr$_{0.12}$CuO$_{4}$ ($T_\mathrm{CDW}$~$\approx$~80\,K) to shed light on this discussion. We observe the emergence of a crystal structure, which is consistent with the CDW modulation in symmetry, well above $T_\mathrm{CDW}$. This global structural change also induces strong fluctuations of local atomic disorder in the intermediate temperature region. At $T_\mathrm{CDW}$, the temperature dependence of this structure develops a kink, while the atomic disorder is minimized. We find that the atomic relaxation dynamics cross over from a cooperative to an incoherent response at $T_\mathrm{CDW}$. These results reveal a rich interplay between the CDWs and atomic fluctuations of distinct spatio-temporal scales. For example, the CDW coherence is enhanced on quasi-elastic timescales by incoherent atomic relaxation.

% However, the underlying phase transition is incomplete, because the atomic disorder induced by it persists down to at least 30\,K. This global structural change also develops a kink at $T_\mathrm{CDW}$, below which the induced atomic disorder is found to be minimized, yet remains finite. %We further study the atomic relaxation dynamics and find a cooperative to incoherent crossover at $T_\mathrm{CDW}$.

%These results highlight that, in addition to having a lattice with minimal disorder, incoherent atomic relaxation plays a critical role in enhancing the CDW coherence on quasi-elastic timescales.
\end{abstract}
\maketitle

Charge density waves (CDWs) are spontaneous modulations of electron density, accompanied by small atomic distortions. They are ubiquitous in the cuprates, where high-$T_\mathrm{c}$ superconductivity arises upon doping with carriers \cite{Comin-ARCMP-2016}.~The phenomenology of CDWs in the cuprates is rich, and complicated.~For example, a well-defined thermodynamic phase transition is missing for the CDWs in most cuprates. Instead, short-range charge correlations prevail at temperatures far exceeding $T_\mathrm{c}$ \cite{Croft-PRB-2014,Miao-NPJQM-2021,Arpaia-Science-2019,Wang-prl-2020}. The widely adopted temperature for characterizing the CDW in the cuprates is $T_\mathrm{CDW}$, below which the in-plane coherence length first increases and eventually saturates at a finite value. Since the CDW strongly intertwines with other electronic degrees of freedom \cite{Keimer-NATURE-2015}, it is important to understand the mechanism defining $T_\mathrm{CDW}$.

Many cuprates undergo a symmetry-breaking structural change on cooling. In the La-based family, there is a high-temperature tetragonal (HTT, space group $\mathit{I4/mmm}$) to low-temperature orthorhombic (LTO, space group $\mathit{Bmab}$) second order phase transition \cite{Yamada-PRB-1998,Axe-prl-1989}. In some compositions, the LTO phase can further transform into a low-temperature tetragonal (LTT, space group $\mathit{P4_\mathrm{2}/ncm}$) one \cite{Axe-prl-1989}. Due to the lack of group-subgroup relationship, this LTO-LTT phase transition is first order in nature, and it can remain incomplete over an exceptionally broad temperature window \cite{Tidey-sr-2022}. Very recently, reflections that are forbidden in the HTT/LTO/LTT space groups have been observed \cite{Sapkota-prb-2021,Frison-prb-2022}. The distorted lattice has been proposed to be monoclinic (space group $\mathit{P2/m}$); this crystal symmetry is consistent with the structure of the CDW observed \cite{Frison-prb-2022}. However, its relationship with $T_\mathrm{CDW}$ is unclear.

% A structural phase transition can induce atomic disorder \cite{Levanyuk1988}. 
In the cuprates, disorder is known to strongly affect the electronic properties \cite{Phillips-rpp-2003}.~The quenched disorder introduced by carrier doping has been studied theoretically \cite{Nie-PNAS-2014}, while experiments have focused on the impact of quenched disorder at a fixed temperature \cite{Campi-nature-2015,Zeljkovic-science-2012}. Reconstructions of local atomic disorder by a phase transition, in contrast, have rarely been explored.

On a disordered lattice, atoms can spontaneously relax \cite{Qiao-jmst-2014}.~Typically, relaxation dynamics are irrelevant for understanding the electronic processes because the latter occur on much faster timescales. However, the CDWs in the cuprates are glass-like \cite{Vojta-AIP-2009,Nie-PNAS-2014}, and fluctuate on exceptionally slow timescales (10$^{-1}$\,s~-~10$^{4}$\,s) \cite{Chen-prl-2016,Mitrovi-prb-2008,Rai-prl-2008,Caplan-prl-2010,Baity-prl-2018}, making the temporally similar atomic relaxation dynamics potentially relevant. 

In this Letter,~we study the atomic lattice in La$_{1.88}$Sr$_{0.12}$CuO$_{4}$, where the in-plane CDW correlation length approximately doubles when the system is cooled below $T_\mathrm{CDW}$~$\approx$~80\,K \cite{Croft-PRB-2014}, using a set of synchrotron x-ray scattering techniques.~In the LTO phase and above $T_\mathrm{CDW}$, symmetry-breaking distortions from which the CDW can evolve by symmetry are seen, along with strong fluctuations of atomic disorder. These local and global structural changes are found to be strongly correlated. At $T_\mathrm{CDW}$, the atomic disorder is at its minimum, while the CDW-compatible distortion gets further promoted. Moreover, the atomic relaxation dynamics below $T_\mathrm{CDW}$ are fundamentally distinct from that above. We discuss the importance of these different atomic fluctuations in understanding the peculiar temperature dependence of the CDW spatial coherence properties.

Two La$_{1.88}$Sr$_{0.12}$CuO$_{4}$ single crystals
% ($\sim$~1.0$\times$1.0$\times$1.0 mm$^{3}$)
were used for this work; they were cut from the same batch grown by the travelling-solvent floating-zone method described in Ref.~\onlinecite{Komiya-prb-2002}.~The hard x-ray diffraction (HXD) experiment was performed on the I16 beamline at the Diamond Light Source~(United Kingdom), using a monochromatic x-ray beam with energy 8.095\,keV. The x-ray diffuse scattering (XDS) and x-ray photon correlation spectroscopy (XPCS) experiment was carried out at the coherent x-ray scattering beamline P10 of the PETRA III storage ring~(Germany).~A monochromatic x-ray beam (energy 8.5\,keV) was focused at the sample position with spot size of 2.5~$\times$~2.5~$\mu$m$^{2}$ in full width at half maximum.~Both experiments were performed in the $\theta$/2$\theta$ reflection geometry.~For simplicity, all Bragg reflections below are indexed using the HTT notation.

\begin{figure}[t]
	\centering
	\includegraphics[width=0.47\textwidth]{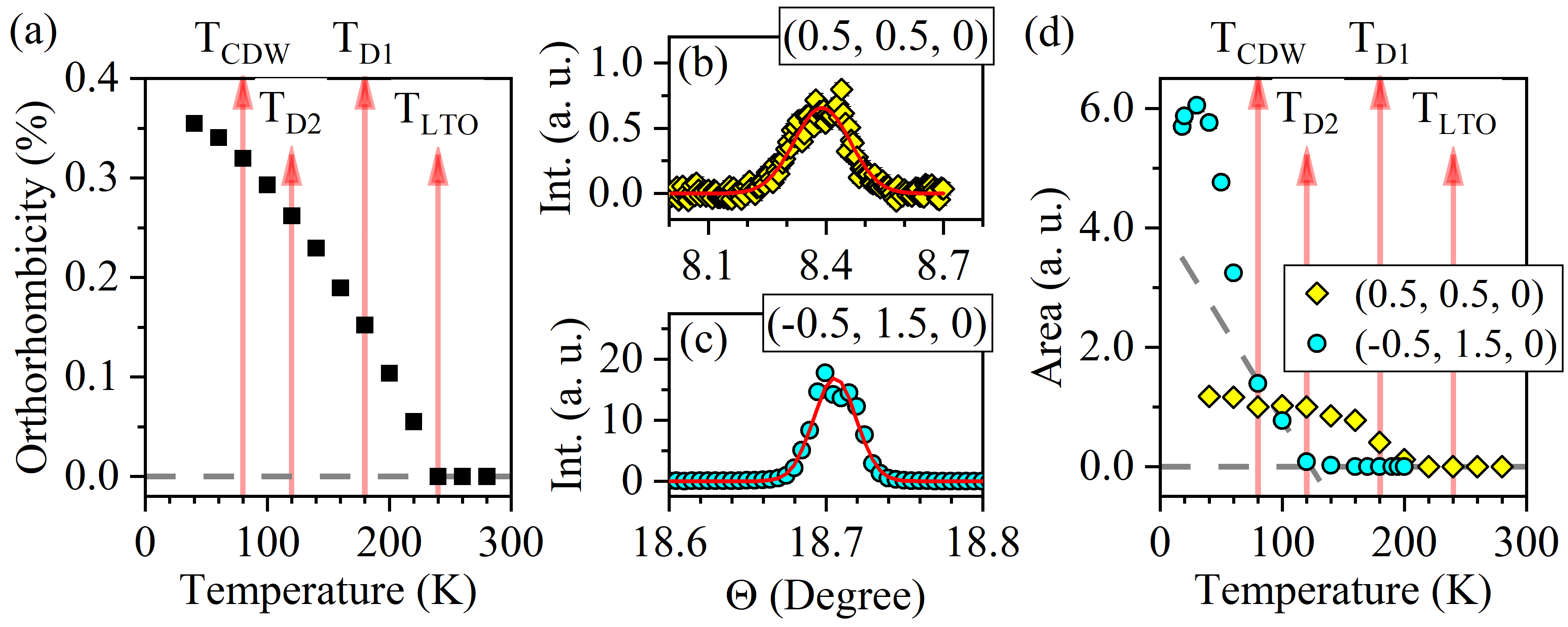}
	\caption{Global atomic distortions. (a) Unit cell orthorhombicity.~(b) and (c) LTO-forbidden reflections at 40\,K. Their areas, obtained by a Gaussian fit (red lines), are shown in (d). Characteristic temperatures are marked by the red arrows (main text). Dashed lines are guides for the eye.
	}
	\label{fig:1}
\end{figure}

The HTT-LTO transition can be characterized by the unit cell orthorhombicity, $|\mathit{a}$-$\mathit{b}|$/$|\mathit{a}$+$\mathit{b}|$, where $\mathit{a}$ and $\mathit{b}$ are the lattice constants. Our HXD measurements show that this occurs at $T_\mathrm{LTO}$~$\simeq$~240~K [Fig.~\ref{fig:1}(a)], in agreement with the literature \cite{Yamada-PRB-1998}. Inside the LTO phase, we observe reflections at (0.5,\,0.5,\,0) [Fig.~\ref{fig:1}(b)] and (-0.5,\,1.5,\,0) [Fig.~\ref{fig:1}(c)], which do not obey the symmetry of its nominal space group $\mathit{Bmab}$ \cite{LS-Notes}. In order to rule out possible higher harmonic contributions from the x-ray source, we have performed careful measurements on some $\mathit{Bmab}$-allowed reflections, e.g.~(1,\,1,\,0). These reflections all appear at $T_\mathrm{LTO}$, with no anomalies at lower temperatures (not shown). Accordingly, we conclude that these $\mathit{Bmab}$-forbidden reflections, which have been reported previously \cite{Frison-prb-2022}, are intrinsic.

We fit their profiles to a Gaussian function.~The peak at (0.5,\,0.5,\,0) is about five times broader than that at (-0.5,\,1.5,\,0). This broadening is primarily due to the in-plane twining induced by the LTO distortion. Specifically, two types of orthogonal twin domains exist below $T_\mathrm{LTO}$ \cite{Horibe-PRB-2000}; this is observed in our HXD measurements too. The $\mathit{Bmab}$-forbidden reflection (0.5,\,-0.5,\,0) scattered by the twin domains, which lies 0.05(1)$^{\circ}$ away from (0.5,\,0.5,\,0) in $\Theta$, causes the peak broadening in Fig.~\ref{fig:1}(b). 

After taking into account the resolution effect, we estimate the coherence length defining the $\mathit{Bmab}$-symmetry-breaking distortions to be at least 400\,$\mathrm{\AA}$. This means that the distortions are non-local, and cannot be attributed to the domain boundaries in the sample \cite{Horibe-PRB-2000}. From a symmetry point of view, these reflections are also forbidden in the LTT space group $\mathit{P42/ncm}$, but they fit the monoclinic space group $\mathit{P2/m}$. %The latter favors the CDW structure 
The CDW structure seen in La$_{1.88}$Sr$_{0.12}$CuO$_{4}$ can develop within this space group \cite{Frison-prb-2022}. In Fig.~\ref{fig:1}(d), we study the temperature dependences of these reflections.~The (0.5,\,0.5,\,0) reflection becomes resolvable at 200\,K, and then increases dramatically at $T_\mathrm{D1}$~=~180\,K. Upon further cooling, another small increase occurs at $T_\mathrm{CDW}$.~The (-0.5,\,1.5,\,0) reflection develops at a lower temperature $T_\mathrm{D2}$~=~120\,K, followed by a sharper rise below $T_\mathrm{CDW}$. Below 30~K, its intensity starts to decrease. One plausible scenario is that the underlying structure gets partially suppressed by the superconductivity, which develops at $T_\mathrm{c}$~$\approx$~30\,K \cite{Yamada-PRB-1998}. This suppression might be linked to the symmetrically compatible CDW, which is known to compete with the superconductivity below $T_\mathrm{c}$ \cite{Croft-PRB-2014}; but our data cannot directly prove it.

% this can be explained by the competition between the CDW and superconductivity \cite{Croft-PRB-2014}, which fits the scenario that the symmetry defining these reflections favors CDW formation \cite{Frison-prb-2022}.

\begin{figure}[b]
	\centering
	\includegraphics[width=0.47\textwidth]{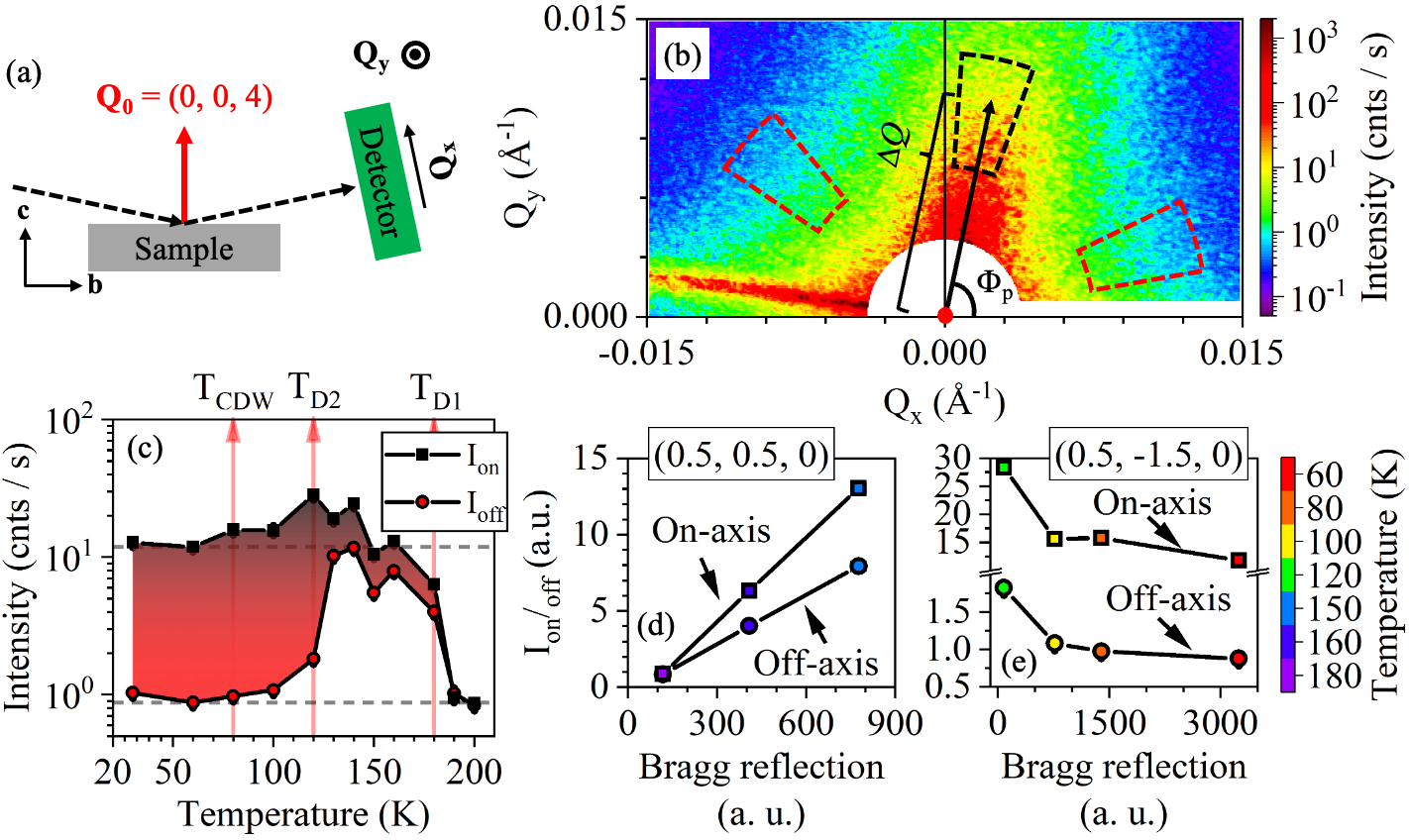}
	\caption{
	Local atomic disorder. (a) XDS/XPCS experimental geometry (top view). (b) XDS around $\mathbf{Q}_\mathrm{0}$~=~(0,\,0,\,4) at 30\,K. A polar coordinate convention centered at $\mathbf{Q}_\mathrm{0}$ (red dot) is defined. Black and red boxes are ensembles used for the data analysis in the main text.~The sharp streak at the bottom-left is a CTR. (c) On- and off- axis XDS vs temperature. (d), (e) Correlations between the area of the Bragg reflection and diffuse scattering intensity. Temperatures are color coded.
	}
	\label{fig:2}
\end{figure}

The HXD measurements indicate that the LTO phase of La$_{1.88}$Sr$_{0.12}$CuO$_{4}$ is not stable at low temperatures. There is a tendency to a further global structural transition, which leads to the LTO-forbidden reflections observed [Fig.~\ref{fig:1}(b)~$\&$~\ref{fig:1}(c)]. Since this type of symmetry breakings are subtle, an accurate structural refinement is challenging. This has been discussed in Ref.~\onlinecite{Frison-prb-2022}, where a monoclinic distortion (space group $\mathit{P2/m}$) is proposed for the low-temperature structure. However, given that the reflection conditions in the LTO lattice are all met by the monoclinic space group, it is not surprising that the average low-temperature crystal structure can be described by the LTO solution (to some extent).

% The HXD measurements reveal additional global atomic distortions which are inconsistent with the LTO symmetry, but which do agree with a crystal structure that can host the CDW structure observed in this compound \cite{Frison-prb-2022}.
We next use XDS to characterize the response of local atomic disorder to these distortions. The data were collected near the $\mathbf{Q}_\mathrm{0}$~=~(0,~0,~4) Bragg reflection. The experimental geometry is shown in Fig.~\ref{fig:2}(a). A polar coordinate system centered at $\mathbf{Q_\mathrm{0}}$ is used to define any relative momentum transfer $\Delta\mathbf{{Q}}$~($\Delta{Q}$,~$\mathit{\Phi}$) = $\mathbf{Q}$ - $\mathbf{Q_\mathrm{0}}$ in the detector plane [Fig.~\ref{fig:2}(b)].~At all temperatures, we observe a sharp crystal truncation rod (CTR) traversing $\mathbf{Q_\mathrm{0}}$ at $\mathit{\Phi}_\mathrm{CTR}$~$\simeq$~172$^{\circ}$. The data in the $\Delta\mathbf{{Q}}$ regions contaminated by the CTR were excluded in the analysis below. 

The XDS pattern at 30~K is shown in Fig.~\ref{fig:2}(b). It is intense and anisotropic, with the principle axis aligned along $\Phi$~=~$\mathrm{\Phi_{p}}$~$\simeq$~78$^{\circ}$. This axis is intact in the temperature window where the XDS is anisotropic. Fundamentally, this XDS anisotropy depends on the local atomic configuration near the defects, and it can be quantitatively understood from the dipole tensor \cite{Lin-PRL-2018}, but exploring this is outside the scope of this Letter, which focuses on the effect of atomic fluctuations on the CDWs.

We analyze the XDS intensities on ($I_{\mathrm{on}}$) and off ($I_{\mathrm{off}}$) the principal axis at $\Delta{Q}$~=~1.07~$\times$~10$^{-2}$~$\mathrm{\AA^{-1}}$. For $I_{\mathrm{on}}$, we sample the ensemble centered at $\mathrm{\Phi_{p}}$ [black boxes in Fig.~\ref{fig:2}(b)]. They have radial and azimuthal coverages of $\pm$0.29~$\times$~10$^{-2}$~$\mathrm{\AA^{-1}}$ and $\pm$10$^{\circ}$, respectively. For $I_{\mathrm{off}}$, in order to avoid the CTR contamination, we sample two ensembles with the same size as those used for $I_{\mathrm{on}}$ but centered at $\mathrm{\Phi_{p}}$~$\pm$~60$^{\circ}$ [red boxes in Fig.~\ref{fig:2}(b)]. Within our resolution, a change in $\Delta{Q}$ or ensemble size does not alter the conclusion made below. 

The results are plotted in Fig.~\ref{fig:2}(c). Both $I_{\mathrm{on}}$ and $I_{\mathrm{off}}$ are small and identical (within the errors) during the initial cooling, meaning that the XDS profile is isotropic. A large signal is seen in both channels at $T_\mathrm{D1}$, accompanied by the development of the $\mathrm{\Phi_{p}}$ anisotropy seen at low temperatures. $\mathrm{I_\mathrm{off}}$ is reentrantly suppressed at $T_\mathrm{D2}$. To further understand the relationship between the global distortions and local disorder, we study the correlations between the area of the LTO-forbidden reflections and $\mathrm{I_\mathrm{on}}$/$\mathrm{I_\mathrm{off}}$. The data around $T_\mathrm{D1}$ and $T_\mathrm{D2}$, where the relevant changes are most dramatic, are plotted in Fig.~\ref{fig:2}(d)~$\&$~\ref{fig:2}(e). We find that the distortion at $T_\mathrm{D1}$ enhances the atomic disorder, while that at $T_\mathrm{D2}$ diminishes it, especially along the off-axis direction. These strong fluctuations of atomic disorder are stabilized at $T_\mathrm{CDW}$, below which $\mathrm{I_\mathrm{on}}$ and $\mathrm{I_\mathrm{off}}$ flatten out [Fig.~\ref{fig:2}(c)].

% The HXD and XDS data indicate that the LTO phase of La$_{1.88}$Sr$_{0.12}$CuO$_{4}$ is not stable at low temperatures. There is a tendency to a further structural transition, which leads to the LTO-forbidden reflections and atomic disorder fluctuations observed [Fig.~\ref{fig:1}(b)~$\&$~\ref{fig:1}(c)]. Since this type of symmetry breakings are subtle, an accurate structural refinement is challenging. This has been discussed in Ref.~\onlinecite{Frison-prb-2022}, where a monoclinic distortion (space group $\mathit{P2/m}$) is proposed. Given that the reflection conditions in the LTO lattice are all met by the monoclinic space group, it is not surprising that the average low-temperature crystal structure can be described by the LTO solution (to some extent).

% A recent study proposed that new structure might be monoclinic \cite{Frison-prb-2022}. Regardless the nature of this new structure, the strong XDS fluctuations between $T_\mathrm{D1}$ and $T_\mathrm{D2}$, together with the persistence of induced atomic disorder down to the lowest temperature probed (30\,K), support the idea that this structural transition is incomplete. 

% Given that the reflection conditions in the LTO lattice are also met in the monoclinic lattice \cite{Frison-prb-2022}, it is not surprising that the average low-temperature crystal structure can be described by the LTO solution (to some extent) in the presence of this incomplete structural change.

\begin{figure}[H]
	\centering
	\includegraphics[width=0.47\textwidth]{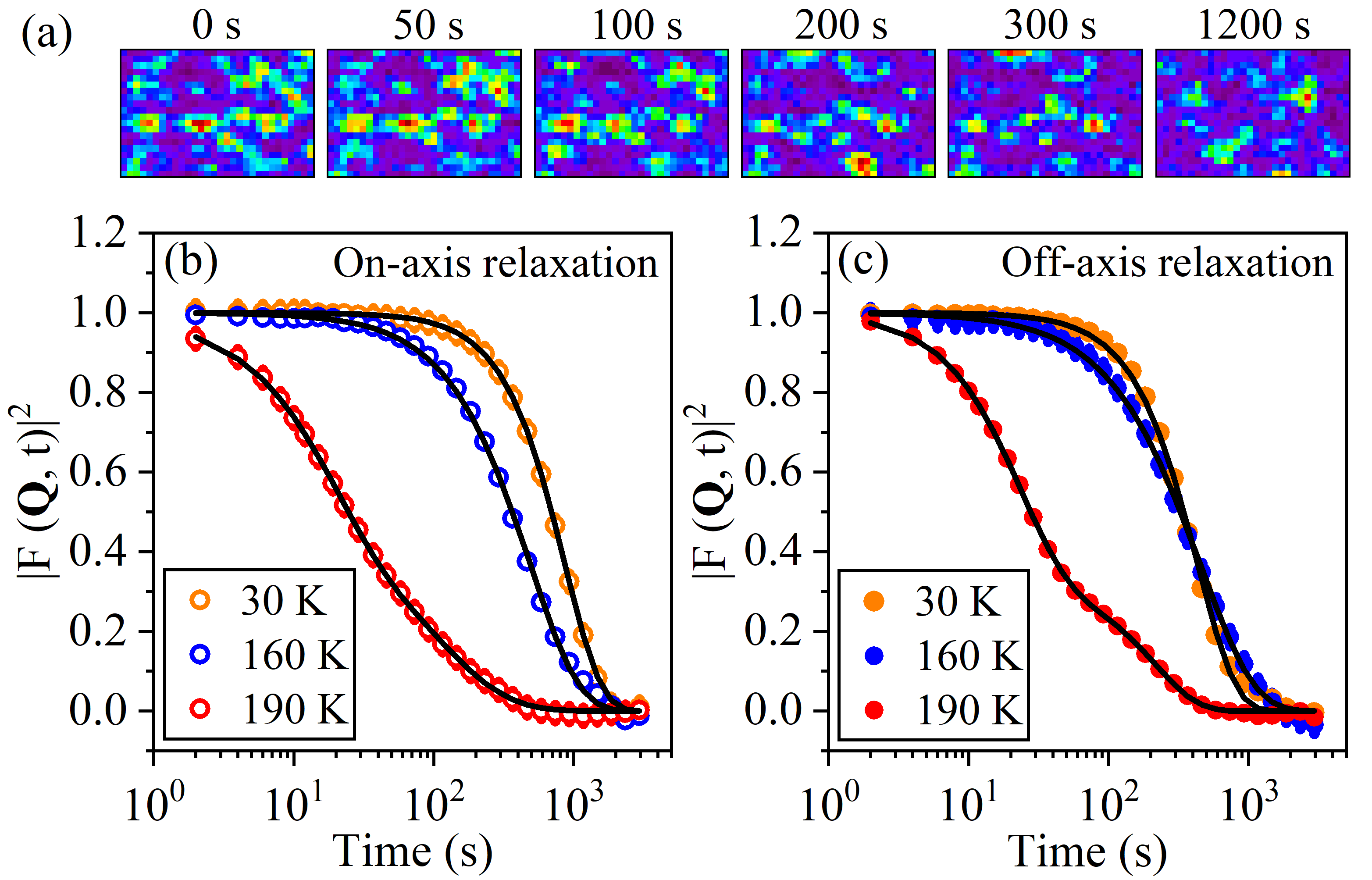}
	\caption{Atomic relaxation. (a) Temporal evolution of speckles in a 20$\times$30 pixel area inside the on-axis ensemble at 160~K. (b) On- and (c) off- axis $|F~(\mathbf{Q},\,t)|^{2}$ vs $\mathit{t}$ at different temperatures.~Solid lines are the single- ($\mathit{T}$~$<$~$T_\mathrm{{D1}}$) or double- ($\mathit{T}$~$\geq$~$T_\mathrm{{D1}}$) KWW fits described in the main text.
	}
	\label{fig:3}
\end{figure}

Having established the complex atomic disorder at low temperatures, which results from the LTO-symmetry-breaking phase transition, we turn to the atomic relaxation dynamics. These can be directly probed by x-ray photon correlation spectroscopy (XPCS) \cite{Caronna-prl-2008,Leitner-nm-2009,ruta-prl-2012,Giordano-nc-2016}, a technique which relies on the configurational change of speckle patterns [Fig.~\ref{fig:3}(a)] -- complex scattering images generated by the interference between the coherent x-rays and local structure --  to calculate the intermediate scattering function $|F~(\mathbf{Q},\,t)|$ in the time ($\mathit{t}$) domain. At each temperature of our experiment, time series of XDS patterns were collected using 1\,s exposure time and a 2\,s interval, up to about 3200\,s. Using these data, we have computed $|F(\mathbf{Q},\,t)|^{2}$ in the same $\mathbf{Q}$ ensembles defined for the XDS characterizations [Fig.~\ref{fig:2}(b)]. The detailed data analysis process for obtaining $|F~(\mathbf{Q},\,t)|^{2}$ is presented in the Supplemental Materials \cite{SuppleM}. 

The $|F(\mathbf{Q},\,t)|^{2}$ data below $T_\mathrm{D1}$, where the primary atomic disorder is related to the structural transition shown in Fig.~\ref{fig:1}(d), are best described by the Kohlrausch–Williams–Watts (KWW) decay model [Fig.~\ref{fig:3}(b) and \ref{fig:3}(c)]:
\begin{equation}
    |F(\mathbf{Q},\,t)|~=~exp\Big[-(\Gamma{}t)^{\beta}\Big].
\end{equation}
In this equation, $\Gamma$ and $\beta$ are the relaxation rate and exponent of the decay process, respectively. The $\beta$ parameter is an indicator for the nature of the dynamics \cite{Madsen-iop-2010,Sinha-AM-2014}. Using Eq.~1, we can quantitatively analyze the atomic relaxation dynamics.

\begin{figure}[t]
	\centering
	\includegraphics[width=0.49\textwidth]{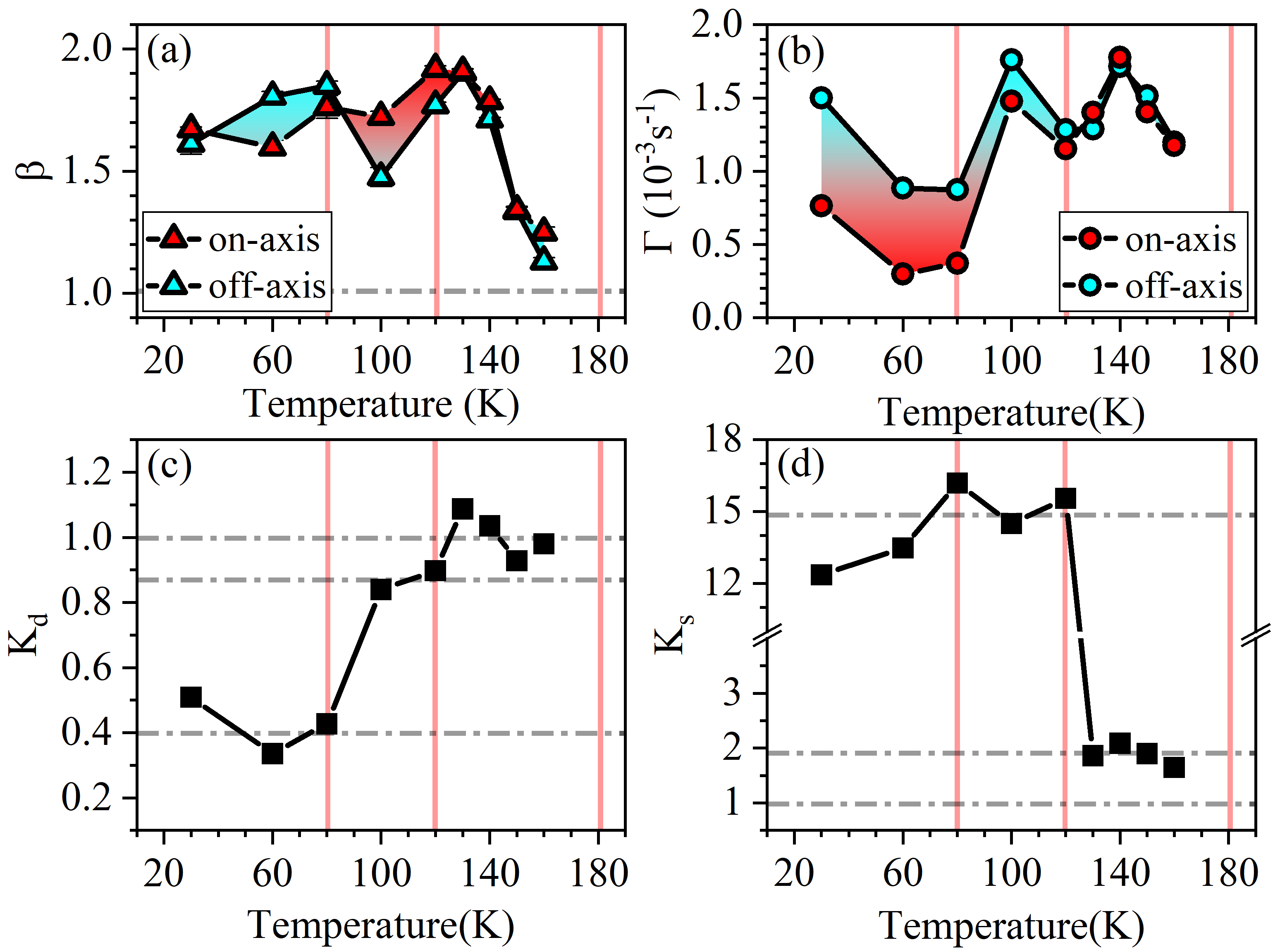}
	\caption{Quantitative analysis of atomic relaxation as a function of temperature. (a) and (b) Relaxation exponent $\beta$ and rate $\Gamma$ on and off the principal XDS anisotropy axis. (c) and (d) Dynamical and static anisotropy parameters. Red vertical lines, from low to high, mark $T_\mathrm{CDW}$, $T_\mathrm{D2}$ and $T_\mathrm{D1}$ (main text). Gray horizontal lines are guides for the eye.}
	\label{fig:4}
\end{figure}

The exponents on ($\beta_\mathrm{on}$) and off ($\beta_\mathrm{off}$) the principal XDS axis are shown in Fig.~\ref{fig:4}(a).~Below $T_\mathrm{{D2}}$, $\beta_\mathrm{on}$ and $\beta_\mathrm{off}$ are larger than 1.5.~On heating from $T_\mathrm{{D2}}$ to $T_\mathrm{{D1}}$, their values decrease. A compressed decay (1.0~$<$~$\beta$~$\leq$~2.0) is commonly observed in glassy systems \cite{Madsen-iop-2010,Sinha-AM-2014,Caronna-prl-2008,ruta-prl-2012,Giordano-nc-2016}. The reductions of $\beta_\mathrm{on}$ and $\beta_\mathrm{off}$ while approaching $T_\mathrm{{D1}}$ are reminiscent of a glass-like transition \cite{Caronna-prl-2008,ruta-prl-2012}. We associate it with the structural transition, or the recovery of the LTO phase, that greatly suppresses the atomic disorder in the sample [Fig.~\ref{fig:2}(c)]. These structural changes are also responsible for the activation of a second KWW decay process when the system is heated across $T_\mathrm{D1}$ [Fig.~\ref{fig:3}(b)~$\&$~\ref{fig:3}(c)]. This phenomenon has been observed in other glassy solids \cite{Giordano-nc-2016}.

The on- ($\Gamma_\mathrm{on}$) and off- ($\Gamma_\mathrm{off}$) axis relaxation rates are shown in Fig.~\ref{fig:4}(b). Their temperature induced variations are random, but non-negligible. This can be explained by the non-equilibrium lattice dynamics in La$_{1.88}$Sr$_{0.12}$CuO$_{4}$ \cite{Horibe-PRB-2000}. For example, using two-time autocorrelation analysis \cite{Sinha-AM-2014,Madsen-iop-2010,SuppleM}, we have seen an aging effect; these results will be reported elsewhere.

We compare the anisotropies of the atomic disorder and relaxation dynamics below $T_\mathrm{{D1}}$, using two unitless parameters: $K_\mathrm{d}$~=~$\Gamma_\mathrm{on}/\Gamma_\mathrm{off}$ and $K_\mathrm{s}$~=~$\mathrm{I_\mathrm{on}}/\mathrm{I_\mathrm{off}}$ [Fig.~\ref{fig:4}(c) and \ref{fig:4}(d)]. Above $T_\mathrm{{D2}}$, the relaxation is almost isotropic ($K_\mathrm{d}$~$\simeq$~1.0), while the atomic disorder already possesses a sizable anisotropy ($K_\mathrm{s}$~$\simeq$~1.9).~In the intermediate temperature region between $T_\mathrm{{CDW}}$ and $T_\mathrm{{D2}}$, no significant change in $K_\mathrm{d}$ can be reliably determined, or at best, it becomes slightly anisotropic ($K_\mathrm{d}$~$\simeq$~0.9). This sharply contrasts with the strongly anisotropic atomic disorder at these temperatures ($K_\mathrm{s}$~$\simeq$~15.0). Profoundly, a sharp anisotropy change in the relaxation dynamics is observed at $T_\mathrm{{CDW}}$, across which $K_\mathrm{d}$~drops to about 0.4 and $K_\mathrm{s}$ only varies smoothly. This anomaly is caused by the slowing down of the relaxation along the principle axis.~As shown in Fig.~\ref{fig:4}(b), the mean value of $\Gamma_\mathrm{on}$ or $\Gamma_\mathrm{off}$ below $T_\mathrm{{CDW}}$ is about 1/3 or 4/5 of that above. 

Atoms can relax either incoherently or cooperatively. In an incoherent process, the dynamics are governed by the local atomic configuration because the atoms are in the non-interacting limit. Accordingly, the relaxation rate $\Gamma$ (Eq.~1) is a monotonically decreasing function of the XDS intensity \cite{dg-physica-1959,Sinha-PBC-1988}. This phenomenon is called de Gennes (dG) narrowing, and it has been observed in diluted glasses \cite{Leitner-nm-2009}. 

Based on the dG narrowing model, $K_\mathrm{d}$ would decrease when the temperature is lowered across $T_\mathrm{{D2}}$, where there is an increase in $K_\mathrm{s}$ [Fig.~\ref{fig:4}(d)]. However, the null or loose coupling between $K_\mathrm{d}$ and $K_\mathrm{s}$ around $T_\mathrm{{D2}}$ does not fit this prediction. Instead, it is strong evidence for cooperative relaxation \cite{Caronna-prl-2008}. Typically, this type of dynamics is driven by inter-atomic interactions. Comparing with its lightly doped counterpart La$_{1.88}$Sr$_{0.08}$CuO$_{4}$, where the dopant (Sr) induced strain field is local \cite{Lin-PRL-2018}, the strain field in La$_{1.88}$Sr$_{0.12}$CuO$_{4}$ is more extensive. The correlation length of the strain field probed in our study can be estimated by 2$\pi$/$|\Delta{Q}|$. As shown in Fig.~\ref{fig:2}(b), the XDS intensity drops by at least two orders of magnitude within a $|\Delta{Q}|$ range of 0.015\,$\mathrm{\AA^{-1}}$, corresponding to a strain field that extends about 500\,$\mathrm{\AA}$ in space. In theory, it is well known that cooperative atomic relaxation can be stabilized by non-local strain fields \cite{Bouchaud-epje-2001,Cipelletti-fd-2003}. Although $K_\mathrm{d}$ does not decrease at $T_\mathrm{{D2}}$, a considerable reduction is observed at $T_\mathrm{{CDW}}$. This supports the idea that the dG narrowing picture, i.e. incoherent atomic relaxation, is recovered below $T_\mathrm{{CDW}}$.

So far, we have shown that the global and local arrangements of atoms in La$_{1.88}$Sr$_{0.08}$CuO$_{4}$, as well as their temporal relaxation dynamics, develop distinct signatures (kinks) at $T_\mathrm{{CDW}}$. These observations have important implications for understanding the CDW coherence properties in this archetypal cuprate system. We summarize them below.

\begin{figure}[b]
	\centering
	\includegraphics[width=0.38\textwidth]{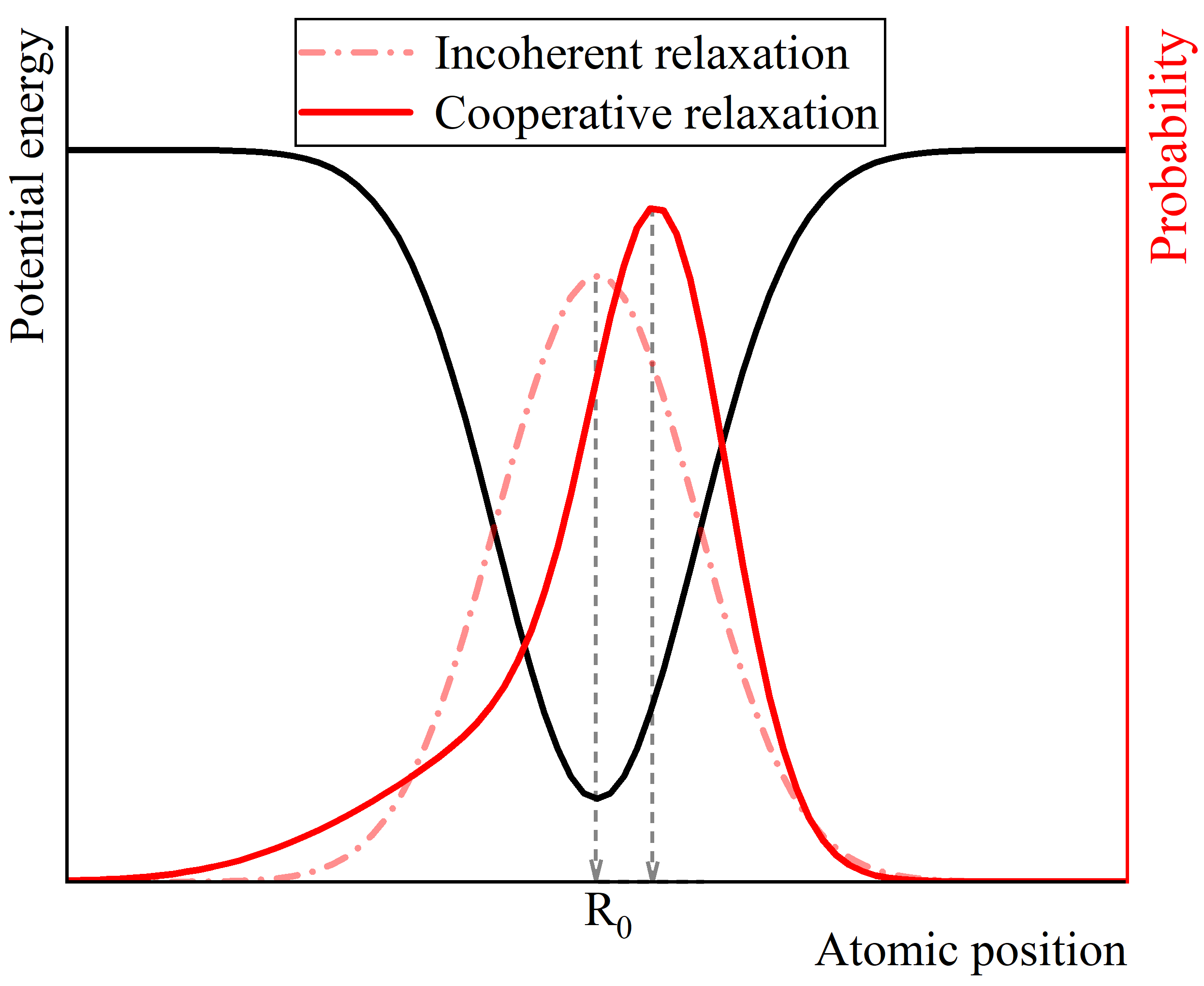}
	\caption{Schematic relaxation of a single atom in real space potential landscape (black solid line). $\mathrm{R_{0}}$ denotes the position of local potential minimum. The probability of finding an atom in space (right axis) maximizes at $\mathrm{R_{0}}$ in an incoherent relaxation process (red dash-dotted line). In a cooperative process, the peak shifts away from $\mathrm{R_{0}}$ (red solid line).}
	\label{fig:5}
\end{figure}

A previous HXD study has reported that quenched disorder can suppress the CDW coherence, leading to a spatially inhomogeneous CDW distribution \cite{Campi-nature-2015}. Our HXD and XDS measurements, on the other hand, reveal that the primary disorder in La$_{1.88}$Sr$_{0.08}$CuO$_{4}$ emerges from a structural phase transition at $T_\mathrm{{D1}}$ [Fig.~\ref{fig:1}(d)]. While the resulting new structure can host the CDW modulation in symmetry \cite{Frison-prb-2022}, the CDWs are insensitive to temperature variations and remain extremely short-ranged in a broad temperature window below $T_\mathrm{{D1}}$ \cite{Croft-PRB-2014,Miao-NPJQM-2021}. We show that this is due to prevailing atomic disorder fluctuations at these temperatures. The CDWs only gain more spatial coherence when the atomic disorder gets maximally suppressed at $T_\mathrm{{CDW}}$ [Fig.~\ref{fig:2}(c)~$\&$~\ref{fig:2}(e)], below which the CDW-compatible structure gets further promoted [Fig.~\ref{fig:1}(d)]. 

Atomic relaxation dynamics have rarely been explored in any cuprate. We have systematically studied it in La$_{1.88}$Sr$_{0.08}$CuO$_{4}$ using XPCS. The measurements suggest that the relaxation dynamics in La$_{1.88}$Sr$_{0.08}$CuO$_{4}$ change from being cooperative above $T_\mathrm{{CDW}}$ to incoherent below.  Since the CDWs in materials of this class fluctuate exceptionally slowly \cite{Chen-prl-2016,Mitrovi-prb-2008,Rai-prl-2008,Caplan-prl-2010,Baity-prl-2018}, atomic relaxation is relevant for determining the CDW coherence on quasi-static timescales. This can be understood based on the real space potential energy landscape for a $\mathit{single}$ atom \cite{Goldstein-jcp-1969}. The CDW-compatible atomic distortion [Fig.~\ref{fig:1}(d)] sets a local potential minimum at $R_0$ (equilibrium position), while the surrounding landscape is controlled by disorder. We use a schematic model to demonstrate the essential processes involved (Fig.~\ref{fig:5}), where the landscape is represented by a Gaussian profile for simplicity. In general, there exists a $R_0$ position for every atom in the distorted lattice. At any given time, the CDW spatial coherence scales with the number of neighboring atoms that can be found $\mathit{simultaneously}$ at their corresponding $R_0$ positions. In the case of incoherent relaxation, the most probable location for finding an atom is $R_0$, because the dynamics are controlled by the local atomic potential (dG narrowing \cite{dg-physica-1959,Sinha-PBC-1988,Leitner-nm-2009}). Cooperative dynamics reconstruct the probability distribution. In other words, $R_0$ is no longer aligned with the optimal probability. As a result, the CDW coherence is suppressed by cooperative relaxation dynamics, which prevail above $T_\mathrm{{CDW}}$ in La$_{1.88}$Sr$_{0.08}$CuO$_{4}$.

Although it is broadly acknowledged that the CDWs in the cuprate superconductors are glass-like \cite{Nie-PNAS-2014,Vojta-AIP-2009} and exhibit quasi-elastic fluctuations \cite{Chen-prl-2016,Mitrovi-prb-2008,Rai-prl-2008,Caplan-prl-2010,Baity-prl-2018}, no existing theory or experiment has accounted for its coupling to the atomic relaxation dynamics which have compatible timescales.~
% Using XPCS, we have studied the atomic relaxation and found that a suppression in the relaxation cooperativeness coincides with the enhancement in the CDW coherence.~Accordingly, we propose a scenario based on the crossover between cooperative and incoherent atomic relaxation dynamics. Using this scenario, we are able to address the peculiar temperature dependence of CDW correlation length below and above $T_\mathrm{{CDW}}$. 
Since almost all cuprate superconductors are intrinsically disordered, and more importantly, possess a significant amount of dopants that can induce non-local strain fields, the coooperative-to-incoherent relaxation dynamics crossover picture described in Fig.~\ref{fig:5} goes beyond the La-based family. As a result, our work opens a new area of studying the many-body interactions in the so far largely unexplored quasi-elastic time regime.

\begin{acknowledgements}
The authors thank Mark Senn and Alessandro Bombardi for fruitful discussions. LS and EB's work is supported in part by Crafoordska stiftelsen (reference number 20190930). This work is also supported by the U.S. Department of Energy, Office of Science, Basic Energy Sciences, Materials Sciences and Engineering Division, under Contract No.~DE-AC02-76SF00515.~JJT acknowledges support from the U.S. DOE, Office of Science, Basic Energy Sciences through the Early Career Research Program. We acknowledge Diamond Light Source for time on Beamline I16 under Proposal MT11098.
\end{acknowledgements}

% \bibliography{lsco}

\begin{thebibliography}{38}
\bibitem [{LS-()}]{LS-Notes}%
  \BibitemOpen
  \href@noop {} {}\bibinfo {note} {In the LTO notation, the (0.5,\,0.5,\,0),
  (0.5,\,-0.5,\,0), (-0.5,\,1.5,\,0) reflections (HTT notation) are re-indexed
  as (1,\,0,\,0), (0,\,1,\,0), and (1,\,2,\,0), respectively; none of these is
  allowed in its space group $\mathit{Bmab}$.}\BibitemShut {Stop}%
\end{thebibliography}


\end{document}